\documentclass[11pt]{article}
\usepackage{cite}
\usepackage{lscape}
\usepackage{rotating}
\usepackage{authblk}

\usepackage{amssymb}
\usepackage{graphicx,wrapfig}
\usepackage{psfrag}
\usepackage{xcolor}
\usepackage{float}
\usepackage{setspace}
\usepackage{booktabs}
\usepackage{multirow}
\usepackage[english]{babel}
\usepackage{marvosym}
\usepackage{tikz}
\usetikzlibrary{arrows,decorations.pathmorphing,backgrounds,positioning,fit,petri,shapes.geometric}
\usetikzlibrary[decorations.markings]
\usetikzlibrary[shapes.arrows]

\definecolor{greytx}{RGB}{70,70,70}

\definecolor{purpletx}{cmyk}{0.4,1,0,0.0}
\definecolor{greentx}{cmyk}{1,0,1.0,0}
\definecolor{mgreentx}{cmyk}{1.0,0,0.5,0}

\def\mang#1;{%
    \begin{scope}[shift={#1}]
        \fill[gray!70] [rounded corners=1.5] (0,0.4) -- (0,0.8) -- (0.4,0.8) -- (0.4,0.4) --
            (0.325,0.4) -- (0.325,0.7) -- (0.3,0.7) -- (0.3,0) -- (0.225,0) --
            (0.225,0.4) -- (0.175,0.4) -- (0.175,0) -- (0.1,0) -- (0.1,0.7) --
            (0.075,0.7) -- (0.075,0.4) -- cycle;
        \fill[gray!70] (0.2,0.92) circle (0.1);
    \end{scope}}

    \def\manc#1;{%
    \begin{scope}[shift={#1}]
        \fill[green!50] [rounded corners=1.5] (0,0.4) -- (0,0.8) -- (0.4,0.8) -- (0.4,0.4) --
            (0.325,0.4) -- (0.325,0.7) -- (0.3,0.7) -- (0.3,0) -- (0.225,0) --
            (0.225,0.4) -- (0.175,0.4) -- (0.175,0) -- (0.1,0) -- (0.1,0.7) --
            (0.075,0.7) -- (0.075,0.4) -- cycle;
        \fill[green!70] (0.2,0.92) circle (0.1);
    \end{scope}}

   \def\manp#1;{%
    \begin{scope}[shift={#1}]
        \fill[violet!80] [rounded corners=1.5] (0,0.4) -- (0,0.8) -- (0.4,0.8) -- (0.4,0.4) --
            (0.325,0.4) -- (0.325,0.7) -- (0.3,0.7) -- (0.3,0) -- (0.225,0) --
            (0.225,0.4) -- (0.175,0.4) -- (0.175,0) -- (0.1,0) -- (0.1,0.7) --
            (0.075,0.7) -- (0.075,0.4) -- cycle;
        \fill[violet!80] (0.2,0.92) circle (0.1);
    \end{scope}}

    \def\womang#1;{%
    \begin{scope}[shift={#1}]
        \fill[gray!70]  [rounded corners=1.0] (0,0.4) -- (0.12,0.8) -- (0.48,0.8) -- (0.6,0.4) --
            (0.525,0.4) -- (0.405,0.7) -- (0.4,0.7) -- (0.5,0.3) -- (0.4,0.3) -- (0.4,0) -- (0.325,0) --
            (0.325,0.3) -- (0.275,0.3) -- (0.275,0) -- (0.2,0) -- (0.2,0.3) -- (0.1,0.3) -- (0.2,0.7) --
            (0.195,0.7) -- (0.075,0.4) -- cycle;
        \fill[gray!70] (0.3,0.92) circle (0.1);
    \end{scope}}

    \def\womanc#1;{%
    \begin{scope}[shift={#1}]
        \fill[green!50]  [rounded corners=1.0] (0,0.4) -- (0.12,0.8) -- (0.48,0.8) -- (0.6,0.4) --
            (0.525,0.4) -- (0.405,0.7) -- (0.4,0.7) -- (0.5,0.3) -- (0.4,0.3) -- (0.4,0) -- (0.325,0) --
            (0.325,0.3) -- (0.275,0.3) -- (0.275,0) -- (0.2,0) -- (0.2,0.3) -- (0.1,0.3) -- (0.2,0.7) --
            (0.195,0.7) -- (0.075,0.4) -- cycle;
        \fill[green!70] (0.3,0.92) circle (0.1);
    \end{scope}}

     \def\womanp#1;{%
    \begin{scope}[shift={#1}]
        \fill[violet!80]  [rounded corners=1.0] (0,0.4) -- (0.12,0.8) -- (0.48,0.8) -- (0.6,0.4) --
            (0.525,0.4) -- (0.405,0.7) -- (0.4,0.7) -- (0.5,0.3) -- (0.4,0.3) -- (0.4,0) -- (0.325,0) --
            (0.325,0.3) -- (0.275,0.3) -- (0.275,0) -- (0.2,0) -- (0.2,0.3) -- (0.1,0.3) -- (0.2,0.7) --
            (0.195,0.7) -- (0.075,0.4) -- cycle;
        \fill[violet!80] (0.3,0.92) circle (0.1);
    \end{scope}}

     \def\manb#1;{%
    \begin{scope}[shift={#1}]
        \fill[cyan!70] [rounded corners=1.5] (0,0.4) -- (0,0.8) -- (0.4,0.8) -- (0.4,0.4) --
            (0.325,0.4) -- (0.325,0.7) -- (0.3,0.7) -- (0.3,0) -- (0.225,0) --
            (0.225,0.4) -- (0.175,0.4) -- (0.175,0) -- (0.1,0) -- (0.1,0.7) --
            (0.075,0.7) -- (0.075,0.4) -- cycle;
        \fill[cyan!70] (0.2,0.92) circle (0.1);
    \end{scope}}

     \def\womanb#1;{%
    \begin{scope}[shift={#1}]
        \fill[cyan!70]  [rounded corners=1.0] (0,0.4) -- (0.12,0.8) -- (0.48,0.8) -- (0.6,0.4) --
            (0.525,0.4) -- (0.405,0.7) -- (0.4,0.7) -- (0.5,0.3) -- (0.4,0.3) -- (0.4,0) -- (0.325,0) --
            (0.325,0.3) -- (0.275,0.3) -- (0.275,0) -- (0.2,0) -- (0.2,0.3) -- (0.1,0.3) -- (0.2,0.7) --
            (0.195,0.7) -- (0.075,0.4) -- cycle;
        \fill[cyan!70] (0.3,0.92) circle (0.1);
    \end{scope}}

         \def\manr#1;{%
    \begin{scope}[shift={#1}]
        \fill[pink!40!red!60] [rounded corners=1.5] (0,0.4) -- (0,0.8) -- (0.4,0.8) -- (0.4,0.4) --
            (0.325,0.4) -- (0.325,0.7) -- (0.3,0.7) -- (0.3,0) -- (0.225,0) --
            (0.225,0.4) -- (0.175,0.4) -- (0.175,0) -- (0.1,0) -- (0.1,0.7) --
            (0.075,0.7) -- (0.075,0.4) -- cycle;
        \fill[pink!40!red!60] (0.2,0.92) circle (0.1);
    \end{scope}}

     \def\womanr#1;{%
    \begin{scope}[shift={#1}]
        \fill[pink!40!red!60]  [rounded corners=1.0] (0,0.4) -- (0.12,0.8) -- (0.48,0.8) -- (0.6,0.4) --
            (0.525,0.4) -- (0.405,0.7) -- (0.4,0.7) -- (0.5,0.3) -- (0.4,0.3) -- (0.4,0) -- (0.325,0) --
            (0.325,0.3) -- (0.275,0.3) -- (0.275,0) -- (0.2,0) -- (0.2,0.3) -- (0.1,0.3) -- (0.2,0.7) --
            (0.195,0.7) -- (0.075,0.4) -- cycle;
        \fill[pink!40!red!60] (0.3,0.92) circle (0.1);
    \end{scope}}

\title{Bridging disconnected networks of first and second lines of biologic therapies in rheumatoid arthritis with registry data: Bayesian evidence synthesis with target trial emulation}
\author[1,*]{Sylwia Bujkiewicz}
\author[1]{Janharpreet Singh}
\author[1]{Lorna Wheaton}
\author[1,2]{David Jenkins}
\author[1]{Reynaldo Martina}
\author[3,4]{Kimme Hyrich}
\author[1,5,6]{Keith R. Abrams}

\affil[1]{Biostatistics Research Group, Department of Health Sciences,
University of Leicester, University Road, Leicester, LE1 7RH, UK}
\affil[*]{sylwia.bujkiewicz@le.ac.uk}
\affil[2]{Centre for Health Informatics, Division of Informatics, Imaging and Data Science, Faculty of Biology, Medicine and Health, University of Manchester, Manchester, UK}
\affil[3]{NIHR Manchester Biomedical Research Centre, Manchester University NHS Foundation Trust, Manchester Academic Health Sciences Centre, Manchester, UK}
\affil[4]{Versus Arthritis Centre for Epidemiology, Centre for Musculoskeletal Research, The University of Manchester, Manchester, M13 9PL, UK}
\affil[5]{Department of Statistics, University of Warwick, , Coventry, CV4 7AL, UK}
\affil[6]{Centre for Health Economics, University of York, York, YO10 5DD, UK}

\begin{document}
\maketitle
\begin{abstract}

Objective:
We aim to utilise real world data in evidence synthesis to optimise an evidence base for the effectiveness of biologic therapies in rheumatoid arthritis in order to allow for evidence on first-line therapies to inform second-line effectiveness estimates.\\
\indent
Study design and setting:
We use data from the British Society for Rheumatology Biologics Register for Rheumatoid Arthritis (BSRBR-RA) to supplement RCT evidence obtained from the literature, by emulating target trials of treatment sequences to estimate treatment effects in each line of therapy. Treatment effects estimates from the target trials inform a bivariate network meta-analysis (NMA) of first and second-line treatments.\\
\indent
Results:
Summary data were obtained from 21 trials of biologic therapies including 2 for second-line treatment and results from six emulated target trials of both treatment lines.  Bivariate NMA resulted in a decrease in uncertainty around the effectiveness estimates of the second-line therapies, when compared to the results of univariate NMA, and allowed for predictions of treatment effects not evaluated in second-line RCTs.\\
\indent
Conclusion:
Bivariate NMA provides effectiveness estimates for all treatments in first- and second-line, including predicted effects in second-line where these estimates did not exist in the data. This novel methodology may have further applications, for example for bridging networks of trials in children and adults.

\end{abstract}

\emph{Keywords: real world evidence, bivariate network meta-analysis, target trial emulation, treatment lines, rheumatoid arthritis, biologic therapies}

\section{Introduction}
The evidence base for health care decision making traditionally consisted of data from randomised controlled trials (RCTs), considered as a gold standard in evaluation of health technologies.
In recent years, there has been growing interest in the use of real world data (RWD) from observational studies in health care evaluation.
Routinely collected data, from electronic health records or patients' registries, can provide useful information about effectiveness of treatments, where data from RCTs may be sparse or are not available at all for some treatment comparisons.
Considerable methodological research has focussed on inclusion of RWD in evidence synthesis with the aim of overcoming some limitations of RCT data \cite{efthimiou2017combining, verde2015combining, welton2020chte2020}.
The focus of such research has been particularly in circumstances where RCT evidence was sparse and combining RCT data with RWD aimed to increase the evidence base to improve the precision of effectiveness estimates \cite{schmitz2013incorporating} and sometimes bridge disconnected networks.

Whilst research to date has largely focussed on exploitation of RWD to mimic or replicate RCT data \cite{bartlett2019feasibility, martina2018inclusion, hernan2016using}, we take a step further to explore use of RWD in a scenario of data generation not typical for the RCT setting.
In this paper, we explored how RWD can be used to optimise an evidence base by using evidence on first-line therapies to inform second-line effectiveness estimates in evidence synthesis.
When data from RCTs are available on effectiveness of a particular treatment, but only in the first line of therapy, a costly trial needs to be undertaken to also evaluate the effectiveness of the new therapy used in patients as a second line treatment (or vice versa).
We investigated the added value of registry data, which provides evidence on both first and second lines in each individual, when amalgamating these data in a network of RCTs for both lines of therapies.
We developed this approach for incorporating RWD into clinical and HTA decision-making using a case study in rheumatoid arthritis (RA).

We made use of data from the British Society for Rheumatology Biologics Register for Rheumatoid Arthritis (BSRBR-RA) to supplement the RCT evidence available only for either first- or the second-line of therapy.
We did so by emulating target trials using the approach developed by Hern$\acute{a}$n and Robins  \cite{hernan2016using}.
We estimated treatment effects of biologic therapies based on the data in emulated target trials, which we then used to inform a bivariate network meta-analysis (NMA) model of first- and second-line treatments.
The estimates from the registry data were used to ``bridge'' disconnected networks for the two lines of therapy. The American College of Rheumatology response criteria (ACR20) was used as an outcome measure.

The remainder of this paper is structured as follows. Data sources and statistical methods are described in Section \ref{sec-methods}. The results are presented in Section \ref{sec-results}, which are followed by discussion and conclusion in Section \ref{sec-discussion}.

\section{Methods}
\label{sec-methods}
\subsection{Data sources}
\subsubsection{Summary data from randomised controlled trials}
Summary data from a literature review of RCTs of biologic therapies in patients with rheumatoid arthritis were obtained for the effectiveness of adalimumab, etanercept, infliximab, golimumab, abatacept and rituximab used as first-line biologic therapies (in biologic naive patients) and the effectiveness of golimumab and rituximab used as second-line biologic therapies in patients who switched from a previous biologic treatment.
Data were obtained from 20 trials including 18 for the first-line treatments and two trials for second-line treatments. 
When constructing a network, placebo arms with methotrexate as concomitant therapy and the arms including a combination of methotrexate and placebo were treated as the same treatment arm in the network.
Methotrexate, used in many trials as part of the combination therapy in the biologic arm, was ignored (for some studies methotrexate was included as concomitant therapy where percentage of patients with addition of methotrexate varied across studies, similarly as in the BSRBR-RA target trials).

\subsubsection{Registry data}
We made use of data from the British Society for Rheumatology Biologics Register for Rheumatoid Arthritis (BSRBR-RA) to supplement randomised trial evidence. Whilst RCTs included only either first- or second-line therapy, registry data provided evidence on both lines of therapy for each patient. BSRBR-RA data consisted of 19410 individuals, 15636 of whom had data recorded on biologic treatment. The data were used to emulate trials of both lines of therapy.

\subsection{Emulation of target trials}
We used the BSRBR-RA data to emulate a series of trials of first- and second-line treatments for a range of biologic therapies using a target trial approach \cite{hernan2016using}.
In the first instance we specified the key components of the target trial protocol, which (following recommendation by Hern$\acute{a}$n and Robins \cite{hernan2016using}) included: eligibility criteria, treatment strategies, assignment procedures, the follow-up period, outcome, causal contrast, and statistical analysis. Note that no RCT included both lines of therapy in sequence, whilst the proposed protocol of the target trial did include the treatment sequence. Therefore, we did not aim for the emulated target trials to resemble any existing RCT (an approach previously used in target trial emulation).

\subsubsection{Eligibility criteria}
Study participants were 18 years of age or older, who had a diagnosis of RA. Patients who were treated with a biologic disease-modifying antirheumatic drug (DMARD) prior to the registration with the BSRBR-RA were excluded.
\subsubsection{Treatment strategies}
Patients had to have received at least two lines of therapy, which could be any of the biologic DMARDs or methotrexate, which is a synthetic DMARD often used as a combination therapy and/or control treatment in trials of biologic therapies in RA patients. Data from patients who switched from first line biologic therapy to no therapy (or to therapies that are neither biologic DMARDs nor methotrexate) were not included.
\subsubsection{Assignment procedures}
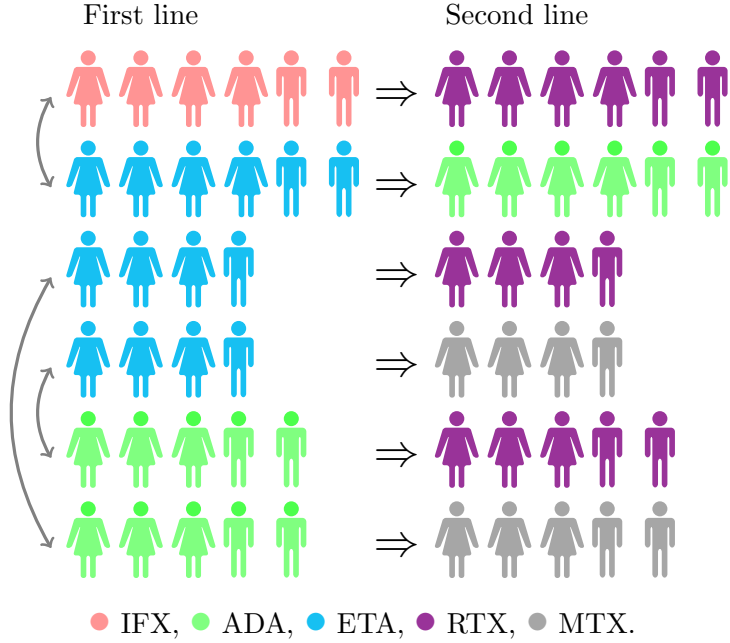
\begin{figure}
\centering
\begin{tikzpicture}
\tiny
\node at(-0.1, 0.4) (arm1) {};
\node at(-0.1, 1.6) (arm2) {};
\node at(-0.1, 2.8) (arm3) {};
\node at(-0.1, 4.0) (arm4) {};
\node at(-0.1, 5.2) (arm5) {};
\node at(-0.1, 6.4) (arm6) {};
\path (arm1.west) edge [<->,very thick, gray, bend left] (arm4.west);
\path (arm5.west) edge [<->,very thick, gray, bend left] (arm6.west);
\path (arm2.west) edge [<->,very thick, gray, bend left] (arm3.west);
\womanc(0,0); \womanc(0.7,0); \womanc(1.4,0); \manc(2.1,0); \manc(2.8,0);
\node at(4.4,0.4) {\LARGE $\Rightarrow$};
\womang(4.9,0); \womang(5.6,0); \womang(6.3,0); \mang(7,0); \mang(7.7,0);
\womanc(0,1.2); \womanc(0.7,1.2); \womanc(1.4,1.2); \manc(2.1,1.2); \manc(2.8,1.2);
\node at(4.4,1.6) {\LARGE $\Rightarrow$};
\womanp(4.9,1.2); \womanp(5.6,1.2); \womanp(6.3,1.2); \manp(7,1.2); \manp(7.7,1.2);
\womanb(0,2.4); \womanb(0.7,2.4); \womanb(1.4,2.4); \manb(2.1,2.4);
\node at(4.4,2.8) {\LARGE $\Rightarrow$};
\womang(4.9,2.4); \womang(5.6,2.4); \womang(6.3,2.4); \mang(7,2.4);
\womanb(0,3.6); \womanb(0.7,3.6); \womanb(1.4,3.6); \manb(2.1,3.6);
\node at(4.4,4.0) {\LARGE $\Rightarrow$};
 \womanp(4.9,3.6); \womanp(5.6,3.6); \womanp(6.3,3.6); \manp(7,3.6);
\womanb(0,4.8); \womanb(0.7,4.8); \womanb(1.4,4.8); \womanb(2.1,4.8); \manb(2.8,4.8);  \manb(3.5,4.8);
\node at(4.4,5.2) {\LARGE $\Rightarrow$};
 \womanc(4.9,4.8); \womanc(5.6,4.8); \womanc(6.3,4.8); \womanc(7,4.8); \manc(7.7,4.8); \manc(8.4,4.8);
\womanr(0,6.0); \womanr(0.7,6.0); \womanr(1.4,6.0); \womanr(2.1,6.0); \manr(2.8,6.0); \manr(3.5,6.0);
\node at(4.4,6.4) {\LARGE $\Rightarrow$};
\womanp(4.9,6.0); \womanp(5.6,6.0); \womanp(6.3,6.0); \womanp(7,6.0); \manp(7.7,6.0);\manp(8.4,6.0);
\node at(1.0,7.5) {\normalsize First line};
\node at(6.0,7.5) {\normalsize Second line};
\end{tikzpicture}\\
\vspace{0.25cm}
\textcolor{pink!40!red!60}{\LARGE$\bullet$} IFX, \textcolor{green!50}{\LARGE$\bullet$} ADA, \textcolor{cyan!70}{\LARGE$\bullet$} ETA, \textcolor{violet!80}{\LARGE$\bullet$} RTX, \textcolor{gray!70}{\LARGE$\bullet$} MTX.
\caption{Schematic diagram representing the process of matching sequence treatment arms. Each row represents patients assigned to unique treatment sequence depicted by different colours. Gray arrows on the left hand side show how the sequence treatment groups were matched; to ensure a balance in terms of the size of each arm and that for both lines the treatment in each arm was different. ADA -- adalimumab, ETA -- etanercept, IFX -- infliximab, RTX -- rituximab, MTX -- methotrexate.}
\label{fig_target_trial}
\end{figure}

Patients were grouped into treatment arms according to the sequence of treatment in two lines of therapy.
These groups of patients (sequence treatment arms) were matched to form experimental and control treatment groups.
Matching was conducted based on size of the trial, ensuring well balanced treatment contrasts, with
methotrexate always taken as the control treatment and rituximab as an experimental treatment. Other biologic therapies could be used as either experimental treatment or control.
The matching procedure had to ensure unique treatments in experimental and control arms for each line. The process is schematically described in Figure \ref{fig_target_trial}.
This procedure resulted in target trials of two lines of therapy recorded on the same patients who switched treatment in both treatment arms.
For example, patients in the experimental arm receiving first line adalimumab switched to infliximab and those in the control arm receiving first line etanercept switched to methotrexate, thus resulting in the first line comparison of adalimumab versus etanercept and in the second line comparison of infliximab versus methotrexate.
Since patients were not randomly allocated, we assumed no unmeasured confounding at baseline conditional on a number of prognostic factors, measured at baseline or initiation of each treatment, that could influence the response. The prognostic factors included age, gender, duration of the disease and a number of clinical measures related to the disease activity including the number of tender and swollen joints, serology (being positive for rheumatoid factor), acute-phase reactants (CRP and ESR) and 28-joint count disease activity score (DAS-28).

\subsubsection{The follow-up period}
The minimum follow up time had to ensure that data were collected 24 weeks after initiation of each line of therapy. Start of the second-line therapy varied depending when patients needed to switch to second-line treatment, which was typically due to either lack of response or adverse reactions.
\subsubsection{Outcome}
Patients were assessed according to ACR20 response criteria, which classified them as responders if they had at least 20\% improvement according to ACR criteria.
Due to a large number of missing values on some of the components of ACR within BSRBR-RA data (the register did not capture patient pain or physician global score), the definition of response was relaxed allowing patients to be classed as responders if they had at least 20\% improvement in at least one of the joint count components (tender or swollen joint count) and at least one of the remaining five components of the ACR measure (physician global assessment, patient global assessment, pain, HAQ, ESR (or CRP)) \cite{felson1995american}.

\subsubsection{Causal contrast and statistical analysis}
Baseline characteristics for each group were summarised to ensure that the covariates were similarly distributed across the treatment arms.
The numbers of responders were then adjusted for covariates using regression adjustment in each line of therapy.
Logistic regression was applied to the data with baseline characteristics and treatment as covariates.
We estimated the per-protocol effect in all emulated trials.

\subsection{Bivariate network meta-analysis}

We used bivariate NMA  to model jointly the treatment effects on ACR20 for first- and second-lines of therapies.
A standard approach to any multivariate meta-analysis is to use a hierarchical model with a multivariate normal distribution used to describe variability at two levels: within-study (where the correlation occurs due to the modelled multivariate quantities, such as treatment effects on multiple outcomes, being measured in the same individuals) and between-studies (where the correlation is a result of heterogeneity between the average effects, measured on each outcome in each study, varying across studies due to, for example, differences in population or treatment doses).
Accounting for the within-study correlation is important in such analysis \cite{riley2009multivariate}.
However, modelling jointly non-normal outcomes, such as Binomial responses, would require transforming data, which can lead to biased results \cite{hamza2008binomial}.
Papanikos et al, carried out a simulation study showing that, when the within-study correlation is weak, a multivariate meta-analysis model with independent binomial likelihoods is preferable \cite{papanikos2020novel}.
Exploratory analysis of the BSRBR-RA data set, estimating the within-study correlation using the bootstrapping approach \cite{daniels1997meta, bujkiewicz2013multivariate}, showed that the within-study correlation between the treatment effects for the two lines of therapy transformed onto the log odds ratio (OR) scale was close to zero.
We, therefore, adapted the approaches to multivariate/bivariate NMA by Achana et al. \cite{achana2014network} and Bujkiewicz et al. \cite{bujkiewicz2019bivariate} by assuming independent binomial likelihoods at the within-study level, as in Papanikos et al \cite{papanikos2020novel}, to model the number of responders $r_{ijk}$ to treatment $k$ in line of therapy $j=1,2$ out of the  number of participants $n_{ik}$ in study $i$, with the probability of response denoted by $p_{ijk}$;

\begin{eqnarray}
r_{ijk} & \sim & Binomial(p_{ijk}, n_{ik}) \nonumber \\
logit(p_{ijk}) & = & \theta_{ijk}\nonumber \\
\theta_{ijk}& = & \left\{
\begin{array}{lll}
\mu_{ijb} & if & k=b \\
\mu_{ijb} + \delta_{ij,bk} & if & k>b
\end{array}
\right.
\nonumber
\end{eqnarray}

\begin{equation}
 \nonumber
\left(
\begin{array}{c}
\delta_{i1,bk}\\
\delta_{i2,bk}
\end{array}
\right) \sim \rm{MVN}
\left(
\left(
\begin{array}{c}
d_{1,bk}\\
d_{2,bk}
\end{array}\right), \;
\mathbf{T}=\left(
\begin{array}{cc}
\tau_{1}^2 & \tau_{1}\tau_{2}\rho\\
\tau_{1}\tau_{2}\rho & \tau_{2}^2
\end{array}
\right)
\right)
\nonumber
\end{equation}
\begin{equation}
 \nonumber
\left(
\begin{array}{c}
d_{1,bk}\\
d_{2,bk}\\
\end{array}
\right)
=
\left(
\begin{array}{c}
d_{1,1k} - d_{1,1b}\\
d_{2,1k} - d_{2,1b}\\
\end{array}
\right)
\label{eq-consistency}
\end{equation}

In this hierarchical model, $\theta_{ijk}$ denotes log odds of response to treatment $k$ in line $j$ and study $i$, $\mu_{ijb}$ is the baseline treatment effect in study $i$ line $j$, and $\delta_{ij,bk}$ are true (random) treatment effects in each study $i$ comparing treatment $k$ with the baseline treatment $b$ in line $j$.
We assume that the true effects follow a bivariate normal distribution that is common to the studies investigating the same treatment contrast $bk$ with mean effect $d_{j,bk}$ and the between-studies covariance matrix $\mathbf{T}$ (here assumed homogeneous across the treatment contrasts).
The network structure of the data is taken into account by assuming that the pooled effects $d_{j,bk}$ for each contrast $bk$ and treatment line $j=1,2$ satisfy the consistency assumption (\ref{eq-consistency}), where $d_{j,1k}$ denote the basic parameters (average treatment effects of each treatment $k=1,\dots, n_t$ in the network relative to the reference treatment $1$ and treatment line $j=1,2$, with $n_t$ denoting the number of treatments in the network.

Prior distributions are placed on the between-studies heterogeneity parameters $\tau_j \sim U(0,2)$,
the between-studies correlation $\rho=r * 2-1$, $r\sim Beta(1.5,1.5)$, the baseline effects $\mu_{ijb}\sim N(0, 10^3)$ and the basic parameters $d_{j,1k}  \sim   N(0, 10^3)$.
When the between-studies correlation is zero, the model reduces to two univariate models for the two outcomes modelled independently; with $\delta_{ij,bk} \sim N((d_{j,1b} - d_{j,1k}), \tau_{j}^2), \; j=1,2$.

To predict treatment effect in the second line when data are only available for the therapy in first line, additional assumptions of exchangeability need to be made, where instead of placing a prior distributions on  basic parameters, we add another level of hierarchy to the model as in Bujkiewicz et al \cite{bujkiewicz2019bivariate}.
For each treatment arm $k$; ancillary parameters $\vartheta_{jk}$  for the two treatment lines $j=1,2$, such that
$d_{j1k}  =  \vartheta_{jk}-\vartheta_{j1}$,
are assumed exchangeable (for the biologic therapies only) and correlated:
\begin{equation}
\left(
\begin{array}{c}
\vartheta_{1k} \\
\vartheta_{2k} \\
\end{array}
\right)
 \sim
N
\left(
\left(
\begin{array}{c}
\eta_1 \\ \eta_2 \\
\end{array}
\right),
\frac{1}{2}
\left(
\begin{array}{cc}
\omega_1^2 & \omega_1 \omega_2 \rho_t \\
\omega_1 \omega_2 \rho_t & \omega_2^2 \\
\end{array}
\right)
\right)
\nonumber
\label{eq-ex}
\end{equation}
for $k=2,\dots, n_t$ (across the biologic therapies only).\\
\vspace{0.25cm}
Prior distributions are placed on the parameters $\omega_j \sim Unif(0,2)$
and $\rho_t =r_t*2-1$ with $r_t \sim Beta(1.5,1.5)$.

\section{Results}
\label{sec-results}
\subsection{Summary of data and the network structure}
Summary data were obtained from 20 RCTs of biologic therapies with 18 trials for first line treatment (including  adalimumab, etanercept, infliximab, golimumab, abatacept and rituximab) and two for second-line treatment (including golimumab and rituximab).
BSRBR-RA data included 12657 individuals given first line biologic at the time of registration.
Follow up data included 112983 observations, which was on average 8.93 follow-ups per individual.
For a large proportion of the visits, methotrexate was recorded as a concomitant therapy to a biologic treatment.
Target trial emulation using the BSRBR-RA data led to generation of six trials of biologic therapies in two lines.
Table \ref{tab-studies} includes information on the number of studies for each treatment contrast, line of therapy and the type of study.

\begin{table}[h]
        \centering
       \begin{tabular}{lcccccc}
       && number && first line && second line\\
source && of studies && treatments && treatments\\
    \hline
\multirow{6}{*}{first line RCTs} &\textcolor{white}{aaaaa}& 6 &\textcolor{white}{aaaaa}& ADA vs MTX &\textcolor{white}{aaaaa}& \\
                                    && 1 && ETA vs MTX && \\
                                    && 4 && IFX vs MTX && \\
                                    && 5 && GOL vs MTX && \\
                                    && 2 && ABT vs MTX && \\
                                    && 1 && RTX vs MTX && \\ \hline
\multirow{3}{*}{second line RCTs} &&   1 && && GOL vs MTX  \\
                                    &&   1 && && RTX vs MTX  \\ \hline
\multirow{10}{*}{target trials}     &&   1 && IFX vs ADA & &RTX vs IFX  \\
                                    &&   1 && ADA vs ETA & &RTX vs IFX \\
                                    &&   1 && ETA vs IFX & &RTX vs ADA \\
                                    &&   1 && IFX vs ETA & &ETA vs MTX \\
                                    &&   1 && ETA vs ADA & &ADA vs MTX \\
                                    &&   1 && ADA vs IFX & &ETA vs MTX \\
                                    \hline
\end{tabular}
\caption{Data sources and treatments; ADA -- adalimumab, ETA -- etanercept, IFX -- infliximab, GOL -- golimumab, ABT -- abatacept, RTX -- rituximab, MTX -- methotrexate.}
\label{tab-studies}
\end{table}

Figure \ref{net-all}a shows the network structure of RCT data for the first and second lines of therapy and Figure \ref{net-all}b illustrates the network structure of target trials emulated from BSRBR-RA data for both lines of therapy.
For the target trials, both treatment lines correspond to the same trial, in contrast to the RCTs which report only either first or second line of treatment.
To emphasise this in Figure \ref{net-all}, we used the same colour of the network edges for both treatment lines for the target trials, in contrast to the RCTs where different colours of edges for different treatment lines represent different trials.
In this paper, we aimed to demonstrate the value of the registry data in estimating the effect of the biologic therapies when used as second-line treatments.
The network of RCT data for the second-line therapy was particularly sparse, including only two trials; for golimumab and rituximab.
BSRBR-RA data gave additional information about adalimumab, etanercept, infliximab as well as rituximab used as second-line therapies.
The network structure for RCT and BSRBR-RA data combined is shown in Figure \ref{net-all}c.

\begin{figure}[p]
\flushleft{a)}\\
\centering
\textcolor{black}{first line} \hspace{4cm} \textcolor{black}{second line}\\
\vspace{0.5cm}
\begin{tikzpicture}[scale=0.7, nodes={draw, circle}, line width=1.5pt]
[inner sep=3mm]
\path node at ( 0,0) [shape=circle,draw=black!40] (pbo) {MTX}
node at ( -3,2) [shape=circle,draw=black!40] (eta) {ETA}
node at ( 0,3) [shape=circle,draw=black!40] (ifx) {IFX}
node at ( 3,2) [shape=circle,draw=black!40] (ada) {ADA}
node at ( -3,-2) [shape=circle,draw=black!40] (gol) {GOL}
node at ( 0,-3) [shape=circle,draw=black!40] (abt) {ABT}
node at ( 3,-2) [shape=circle,draw=black!40] (rtx) {RTX}
(ada) edge[violet!80]  (pbo)
(eta) edge[violet!80]  (pbo)
(ifx) edge[violet!80]  (pbo)
(gol) edge[violet!80]  (pbo)
(abt) edge[violet!80]  (pbo)
(rtx) edge[violet!80]  (pbo)
node at ( 8,0) [shape=circle,draw=black!40] (pbo2) {MTX}
node at ( 5,-2) [shape=circle,draw=black!40] (gol2) {GOL}
node at ( 8,-3) [shape=circle,draw=white] (abt2) {\textcolor[rgb]{1.00,1.00,1.00}{ABT}}
node at ( 11,-2) [shape=circle,draw=black!40] (rtx2) {RTX}
(gol2) edge[green!90]  (pbo2)
(abt2) edge[white]  (pbo2)
(rtx2) edge[green!90]  (pbo2);
\end{tikzpicture}\\
\flushleft{b)}\\
\centering
\textcolor{black}{first line} \hspace{4cm} \textcolor{black}{second line}\\
\vspace{0.5cm}
\begin{tikzpicture}[scale=0.7, nodes={draw, circle}, line width=1.5pt]
[inner sep=3mm]
\path
node at ( -3,2) [shape=circle,draw=black!40] (eta) {ETA}
node at ( 0,3) [shape=circle,draw=black!40] (ifx) {IFX}
node at ( 3,2) [shape=circle,draw=black!40] (ada) {ADA}
(ada) edge[cyan!80]  (eta)
(eta) edge[cyan!80]  (ifx)
(ifx) edge[cyan!80]  (ada)
node at ( 8,0) [shape=circle,draw=black!40] (pbo2) {MTX}
node at ( 5,2) [shape=circle,draw=black!40] (eta2) {ETA}
node at ( 8,3) [shape=circle,draw=black!40] (ifx2) {IFX}
node at ( 11,2) [shape=circle,draw=black!40] (ada2) {ADA}
node at ( 11,-2) [shape=circle,draw=black!40] (rtx2) {RTX}
(ada2) edge[cyan!80]  (pbo2)
(eta2) edge[cyan!80]  (pbo2)
(rtx2) edge[cyan!80]  (ifx2)
(rtx2) edge[cyan!80]  (ada2);
\end{tikzpicture}\\
\flushleft{c)} \\
\centering
\textcolor{black}{first line} \hspace{4cm} \textcolor{black}{second line}\\
\vspace{0.5cm}
\begin{tikzpicture}[scale=0.7, nodes={draw, circle}, line width=1.5pt]
[inner sep=3mm]
\path
node at ( 0,0) [shape=circle,draw=black!40] (pbo) {MTX}
node at ( -3,2) [shape=circle,draw=black!40] (eta) {ETA}
node at ( 0,3) [shape=circle,draw=black!40] (ifx) {IFX}
node at ( 3,2) [shape=circle,draw=black!40] (ada) {ADA}
node at ( -3,-2) [shape=circle,draw=black!40] (gol) {GOL}
node at ( 0,-3) [shape=circle,draw=black!40] (abt) {ABT}
node at ( 3,-2) [shape=circle,draw=black!40] (rtx) {RTX}
(ada) edge[violet!80]  (pbo)
(eta) edge[violet!80]  (pbo)
(ifx) edge[violet!80]  (pbo)
(gol) edge[violet!80]  (pbo)
(abt) edge[violet!80]  (pbo)
(rtx) edge[violet!80]  (pbo)
(ada) edge[cyan!80]  (eta)
(eta) edge[cyan!80]  (ifx)
(ifx) edge[cyan!80]  (ada)
node at ( 8,0) [shape=circle,draw=black!40] (pbo2) {MTX}
node at ( 5,2) [shape=circle,draw=black!40] (eta2ifx2) {ETA}
node at ( 8,3) [shape=circle,draw=black!40] (ifx2) {IFX}
node at ( 11,2) [shape=circle,draw=black!40] (ada2) {ADA}
node at ( 5,-2) [shape=circle,draw=black!40] (gol2) {GOL}
node at ( 8,-3) [shape=circle,draw=white] (abt2) {\textcolor[rgb]{1.00,1.00,1.00}{ABT}}
node at ( 11,-2) [shape=circle,draw=black!40] (rtx2) {RTX}
(ada2) edge[cyan!80]  (pbo2)
(eta2) edge[cyan!80]  (pbo2)
(ifx2) edge[cyan!80]  (pbo2)
(rtx2) edge[cyan!80]  (ada2)
(gol2) edge[green!90]  (pbo2)
(rtx2) edge[green!90]  (pbo2)
(abt2) edge[white]  (pbo2)
(rtx2) edge[cyan!80]  (ifx2);
\end{tikzpicture}
\caption{\large Network diagram for (a) the RCT data, (b) BSRBR-RA data and (c) combined data; first line treatments (left) and second line treatments (right).}
\label{net-all}
\end{figure}

\subsection{Results of network meta-analyses}
Results of two univariate NMAs  of biologic therapies used as second-line treatments are shown in Table \ref{tab-unma}, where the  lower triangle includes the results from the NMA using data from RCTs alone and the upper triangle shows results of NMA combining data from RCTs and BSRBR-RA register.
Including the registry data allowed for estimation of treatment effects for second line biologic therapies which were not included in the RCT network.
There was also some improvement in the precision of treatment effect estimates for those already included in the RCT data.
For example, comparing rituximab with methotrexate gave OR=11.34 (95\% CrI: 0.37, 59.4) when using RCT data alone, whilst addition of the registry data resulted in OR=3.81 (0.48, 14.5) thus reducing the uncertainty by 76\% in terms of the width of the credible interval. The between-studies correlation was weak; $\rho$=-0.17 (95\% CrI: -0.86, 0.8).

\begin{table}
\centering
       \begin{tabular}{l|c|c|c|c|c|c|c}
	&	MTX	&	ADA	     &	     ETA	  &	   IFX	&	GOL	&	ABT	&	RTX	\\ \hline
	&		&	4.69 	 &	4.41 	  &	11.0 	&	4.89 	&		&	3.81 	\\
MTX	&		&(0.6, 17.5) &(0.79, 14.5)	& (0.56, 50.9)	& (0.32, 21.5)	&	--	&	(0.48, 14.5)	\\ \hline
	&		&		     &	2.14 	  &	        3.42 	&	2.66 	&		&	1.18 	\\
ADA	&	--	&		     &	(0.11, 9.67)	&	 (0.17, 15.5)	& (0.05, 12.1)	&	--	& (0.15, 4.41)\\ \hline
	&		&		     &		     &	4.90  	&	2.17 	&		&	1.60 	\\
ETA	&	--	&	--	      &		     &	 (0.11, 23.1) 	& (0.06, 10.1)	&	--	&	 (0.08, 7.36)	\\ \hline
	&		&		     &		     &		&	2.76 	&		&	0.66 	\\
IFX	&	--	&	--	      &	--	       &		&	 (0.02, 10.8)	&	--	&	 (0.11, 2.19)	\\ \hline
	&	6.41	&		&		      &		&		&		&	3.31 	\\
GOL	&	(0.21, 33.4)	&	--	        &	--	&	--	&		&	--	&	 (0.07, 15.0)	\\ \hline
	&	    	&			&		&		& 	&		&		\\
ABT	&	    --	&		--	&	--	&	--	& --	&	\textcolor[rgb]{1.00,1.00,1.00}{(0.10, 00.1)} &	--	\\ \hline
	&	11.34 	&		&		&		&	15.7 	&		&		\\
RTX	&	(0.37, 59.4)	&	--	&	--	&	--	&	 (0.05, 64.3)	&	-- 	&		\\
\end{tabular}
\caption{\label{tab-unma}Results of a  univariate NMA  of biologic therapies used as second-line treatments
 using data from RCTs alone (lower triangle) and combining data from RCTs and BSRBR-RA registry (upper triangle).}
\end{table}

The results of a  bivariate NMA combining data from RCTs and BSRBR-RA of biologic therapies in both lines of therapy are shown in Table \ref{tab-bnmas}, with the results  using the ``standard'' bivariate NMA model in the upper triangle
and the results from the analysis assuming exchangeability of biologic therapies in the lower triangle.
Combining data from the first and second lines of therapy through the use of the bivariate NMA led to a further  decrease in uncertainty for most of the individual treatments when compared to the results of the univariate NMA of second-line therapy alone.
For example, ACR20 response to adalimumab vs methotrexate from the bivariate NMA was OR=4.36 (0.67, 15.5) compared to OR=4.69 (0.6, 17.5) from the univariate NMA.

The bivariate NMA approach assuming the additional exchangeability of the absolute effects of the biologic therapies allowed for predictions of treatment effects that had not been evaluated in trials in a second-line setting. In this case, it produced effectiveness estimates for abatacept in the second line of therapy against all other treatments in the network.
Moreover, this additional exchangeability led to a noticeable reduction in uncertainty around the remaining estimates of effect for other therapies. This was a result of additional borrowing of information.
However, there may have been some degree of smoothing of the effects across the biologic therapies, which was difficult to judge due to the large uncertainty. A sensitivity analysis was carried out using a \emph{t}-distribution in place of the normal distribution in model (\ref{eq-ex}), which largely produced very similar results but inflated the uncertainty around the effectiveness estimate for abatacept in the second line.

\begin{table}
\centering
\begin{tabular}{l|c|c|c|c|c|c|c}
	&	MTX	     &	ADA	      &	ETA	&	IFX	&	GOL	&	ABT	&	RTX	\\ \hline
MTX	&		     &	4.36 	    &	4.03 	&	      10.2 	&	      5.07 	&		&	3.54 	\\
	&		        & (0.67, 15.5)	&(0.86, 12.3)& (0.53, 46.7)	&	(0.3, 23.0)	&	--	&	(0.44, 13.5)\\ \hline
	&	3.33	&		     &	1.91 	        &	3.0 	&        	2.51 	&		&	1.06 	\\
ADA	&	(1.37, 6.97)&		&	(0.12, 8.22)	& (0.18, 13.3)	& (0.06, 12.1)	&	--	& (0.16, 3.75)	\\ \hline
	&	3.36	  &	 1.12	&		&	    4.63&	              2.12 	&		&	1.50 	\\
ETA	&	(1.5, 6.58)& (0.44, 2.41)	&		& (0.11, 21.7) 	& (0.06, 10.4)	&	--	& (0.08, 6.87)	\\ \hline
	&	4.2	     &	1.34	&	        1.33	&		   &       	2.89 	&		&	0.62 	\\
IFX	&	(1.42, 10.6)& (0.53, 3.26)	& (0.48, 3.43)	&		& (0.02, 12.1)	&	--	&	 (0.12, 1.96)	\\ \hline
	&	3.28  	&	   1.07	&	         1.05	&	        0.91	&		&		&	3.18 	\\
GOL	&	(1.16, 7.38)& (0.36, 2.45)	& (0.35, 2.36)	&	(0.24, 2.03)&		&	--	&	 (0.06, 14.6)	\\ \hline
	&	  4.08  	&	1.31	&	     1.29	&	     1.07	&  1.42	&		&		\\
ABT	&(0.77, 12.3)&	(0.26, 3.87)&	(0.24, 3.76)&	(0.18, 2.92)& (0.26, 4.43)	&	 &	--	\\\hline
	&	2.82 	&	0.89	&	      0.9	&	     0.75	&	0.97 	&	         1.1	&		\\
RTX	&(1.09, 5.96) &(0.39, 1.65)	&	(0.32, 1.8)	&	(0.3, 1.31) & (0.32, 2.19)	&	(0.21, 2.96) 	&		\\
\end{tabular}
\caption{\label{tab-bnmas}Results of a  bivariate NMA combining data from RCTs and BSRBR-RA of biologic in both lines of therapy using the ``standard'' bivariate NMA model (upper triangle) and assuming exchangeability of biologic therapies (lower triangle).}
\end{table}

\section{Discussion and conclusions}
\label{sec-discussion}

In this paper, we provide a conceptual approach for using RWD, such as from registries or electronic health records, to generate estimates of effectiveness of treatments in first and second lines of therapy and combining them with RCT data to enhance the evidence base and provide effectiveness estimates of therapies in the second line, where data on effectiveness in the second line are not available from RCTs.
In such circumstances, producing these estimates would require conducting expensive and time-consuming additional clinical trials.
The proposed approach can be used to carry out feasibility analysis or provide inputs to the trial design or even be used for evidence based decision-making where evidence is sufficiently robust.

When carrying out this research, we came across a number of limitations.
Some of them were related to data.
In particular, the RCT data were relatively sparse with a star-shaped network for the first line treatments and only two trials reporting the effectiveness of biologic therapies in the second line.
The data set was simplified by combining the control arms (including methotrexate as either combination therapy or concomitant therapy with placebo) into the same control arm denoted as methotrexate. This was done to strengthen the network structure to better illustrate the methodological aspect of this work.
Most of the biologic arms also included methotrexate.
Considering that for a large proportion of visits in the BSRBR-RA data methotrexate was recorded as a concomitant therapy to a biologic treatment, an assumption was made that a large proportion of patients receiving biologic therapy, across all studies, also received methotrexate.
The registry data set contained a substantial amount of missing data, in particular for some of the components of the ACR20 response criteria, which was not due to the issues of quality of the data but owing to the fact that some of the components are not routinely collected by the register.
In order to estimate the response to the biologic therapies, we chose to relax the definition of the response.
In addition, the register only captures 28 joint counts, which may be different from some of the trials.
Considering these potentially strong assumptions around the data sources, the results of our analysis should not be used for clinical interpretation, but only as an illustration of the proposed methodology.

There were  only six target trials generated from the registry data, which resulted in   substantial uncertainty around the between-studies correlation, as these were the only studies contributing data to estimating the correlation.
The combined network was still limited with lack of data on each contrast and line across study designs.
Target trial data were incorporated in the NMAs at face value, assuming they were equivalent with RCT data. Extensions of the analysis could include a power prior approach \cite{ibrahim2000power}, allowing for down-weighting of RWD, or hierarchical modelling to differentiate between the two study designs \cite{schmitz2013incorporating}.
Further investigation into data scenarios and model assumptions needs to be carried  out to understand when this framework can be most efficient.

\subsection{Conclusions}
Registry data can be used to bridge networks of first and second lines of therapy which are disconnected when using RCT data alone.
Bivariate NMA of combined data from RCTs and RWD can be used to predict effectiveness of a treatment in second line use when the therapy is only investigated in a RCT as first line (or vice versa).
The approach can be applied to other settings where RCT data are available for disjoint subsets of population, such as, for example, children and adults and registries may provide data covering follow-up period from childhood to adulthood for each individual.

\section{Contributions}
SB and KRA conceived the research idea. SB conceptualised the study. SB, JS, LW, DJ and KH curated the data. SB and JS undertook formal analyses.  KH provided clinical input. SB and KRA provided supervision for JS, LW and DJ. SB contributed the original draft of the manuscript and visualisation. All authors contributed to manuscript revisions.

\section{Acknowledgements}
The early work leading to these results has received support from the Innovative Medicines Initiative Joint Undertaking under grant agreement n$^o$ [115546], resources of which are composed of financial contribution from the European Union’s Seventh Framework Programme (FP7/2007- 2013) and EFPIA companies’ in kind contribution.
In subsequent research KRA and SB were supported by the UK Medical Research Council (MRC) Methodology Research Panel (MRP) [grant no. MR/R025223/1].
SB was also partially supported by the MRC MRP [MR/L009854/1].
JS and LW were supported by UK National Institute for Health Research (NIHR) Methods Fellowship [award no. RM-FI-2017-08-027].
KLH is supported by the NIHR Manchester Biomedical Research Centre.

The authors would like to thank the British Society for Rheumatology for use of this data. The BSR Biologics Register in Rheumatoid Arthritis (BSRBR-RA) is a project owned and run by the BSR on behalf of its members. BSR would like to thanks its funders for their support, currently AbbVie, Amgen, Celltrion, Eli Lilly, Pfizer, Samsung, Sandoz, Sanofi, and in the past Roche, UCB, SOBI and Merck. This income finances a wholly separate contract between BSR and The University of Manchester to host the BSRBR-RA. All decisions concerning analyses, interpretation and publication are made autonomously of any industrial contribution. The BSRBR-RA would like to gratefully acknowledge the support of BSR members and the National Institute for Health Research, through the Comprehensive Local Research Networks at participating centres.

\section{Disclosures}
SB has served as a paid consultant, providing methodological advice, to NICE, Roche and RTI Health Solutions, has received payments for educational events from Roche and has received research funding from European Federation of Pharmaceutical Industries \& Associations (EFPIA) and Johnson \& Johnson.
JS,LW,DJ do not have conflict of interest.
RM has served as a consultant for Roche,Takeda, Freeline and Khondrion
KH received honoraria from Abbvie (Speaker fees) and grant income from Pfizer and BMS.
KRA has served as a paid consultant, providing methodological advice, to;
Abbvie, Amaris, Allergan, Astellas, AstraZeneca, Boehringer Ingelheim,
Bristol-Meyers Squibb, Creativ-Ceutical, GSK, ICON/Oxford Outcomes, Ipsen,
Janssen, Eli Lilly, Merck, NICE, Novartis, NovoNordisk, Pfizer, PRMA, Roche and
Takeda, and has received research funding from Association of the British
Pharmaceutical Industry (ABPI), European Federation of Pharmaceutical
Industries \& Associations (EFPIA), Pfizer, Sanofi and Swiss Precision Diagnostics.
He is a Partner and Director of Visible Analytics Limited, a healthcare
consultancy company.

\section{References}
\bibliographystyle{unsrt}
\bibliography{ra_lines_refs}

\end{document}